\documentclass[copyright,creativecommons]{eptcs}

\usepackage{parskip,etex}
\usepackage{hyperref}
\usepackage{xspace}
\providecommand{\event}{GANDALF'14} 
\usepackage[beamer,thm,fleqn,nolmodern]{sk2}
\usepackage[]{tikz,tikz-qtree}
\usetikzlibrary{arrows,automata,shadows,matrix,shapes,calc}

\usepackage{macros}

\title{Deterministic Automata for Unordered Trees}
\author{Adrien Boiret%
\thanks{University of Lille 1, France}
\thanks{Links (Inria Lille \& LIFL, UMR CNRS 8022), France}
\and
Vincent Hugot
\thanks{Inria}
\footnotemark[2]
\and
Joachim Niehren\footnotemark[3]\,\,\footnotemark[2]
\and Ralf Treinen\thanks{University Paris Diderot, PPS (UMR CNRS 7126), France}
}
\def\titlerunning{Deterministic Automata for Unordered Trees}
\def\authorrunning{Boiret, Hugot, Niehren \& Treinen}

\ifdefined\CMPused\else\def\CMPshort{}\fi

\usepackage[long,hhmmss]{datetime}
\newcommand{\compileMark}{Compile time: \bf\today\ at \currenttime.}

\usepackage[%
] {comments}

\defineCommentingMacros  {Vincent}   {purple}    {vh}
\defineCommentingMacros  {Adrien}    {blue}      {ab}
\defineCommentingMacros  {Joachim}   {green}     {jn}
\defineCommentingMacros  {Ralf}      {orange}    {rt}

\newif\ifLongPaper
\LongPapertrue  

\makeatletter 
\ifdefined\CMPshort \LongPaperfalse\@comments@display@false \fi
\ifdefined\CMPlong  \LongPapertrue \@comments@display@false \fi
\ifdefined\CMPshortC \LongPaperfalse\@comments@display@true \fi
\ifdefined\CMPlongC  \LongPapertrue \@comments@display@true \fi
\makeatother 

\begin{document}
\maketitle

\begin{abstract}

Automata for unordered unranked trees are relevant for defining schemas
and queries for data trees in \JSON or \XML format. While
the existing notions are well-investigated concerning
expressiveness, they all lack a proper notion of determinism, which 
makes it difficult to distinguish subclasses of
automata for which problems such as inclusion, equivalence, and minimization 
can be solved efficiently. In this paper, we propose and investigate
different notions of ``horizontal determinism'', starting from automata for
unranked trees in which the horizontal
evaluation is performed by finite state automata. We show that a 
restriction to confluent horizontal evaluation leads to 
polynomial-time emptiness and universality, but still suffers from
coNP-completeness of the emptiness of binary intersections. Finally, 
efficient algorithms can be obtained by imposing an order of horizontal 
evaluation globally for all automata in the class. Depending on
the choice of the order, we obtain different classes of automata, each
of which has the same expressiveness as Counting \MSO. %
\vhm{\compileMark}%

\end{abstract}

\section{Introduction}

Logics and automata for unordered trees were studied in the last twenty years mostly for querying \XML documents \cite{SeidlSchwentickMuscholl03,BonevaTalbot05,DalZilioLugiez03} and more recently for querying \NoSQL databases \cite{BenzakenCastagnaNguyen13}. They were already studied earlier, for modeling feature structures
in computational linguistics \cite{Smolka92} and records in programming 
languages \cite{SmolkaTreinen94,MuellerNiehrenTreinen98,NiehrenPodelski93}.

\begin{figure}[t]
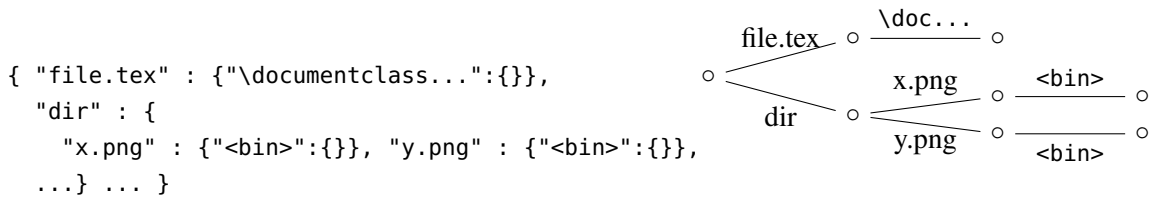

\begin{minipage}[t]{9cm}
\begin{verbatim}
{ "file.tex" : {"\documentclass...":{}},
  "dir" : {
    "x.png" : {"<bin>":{}}, "y.png" : {"<bin>":{}},
  ...} ... }
\end{verbatim}
\end{minipage}
\begin{minipage}[b]{5cm}
\[
\tzt [ht,ps,loose]{\circ}{
   \edge node [above] {\x{file.tex}}; [.{\circ}
     \edge node [above] {\texttt{\textbackslash doc...}};{\circ} ]
   \edge node [below] {\x{dir}}; [.{\circ}
     \edge node [above] {\x{x.png}};[.{\circ} \edge node [above] {\texttt{<bin>}};{\circ} ]
     \edge node [below] {\x{y.png}};[.{\circ} \edge node [below] {\texttt{<bin>}};{\circ} ]
    ]
     }
\]
\end{minipage}
\caption{\label{json:intro}Unordered trees in \JSON format, describing a typical file system.}
\end{figure}

In this paper, we shall consider unordered unranked data trees
whose edges are labeled with strings over a finite alphabet, 
so that there are infinitely many such data values. 
For instance, we can consider a directory of a Linux file system as an unordered tree
(when ignoring 
symbolic links and multiple hard links to files) given in \JSON (the JavaScript Object
Notation \cite{Nurseitov09}), as for instance in Figure \ref{json:intro}. 
This is a recent language-independent format for nested key-value stores,
which already found much interest in Web browsers and for \NoSQL databases
such \textsc{Ibm}'s \textsc{Jaql} \cite{Beyer11}. 
In this representation, we might want to verify that a \LaTeX\xspace repository contains
exactly one main file, i.e. at most one file matching \texttt{*".tex"} whose
content matches \Document. This property can be checked by formul\ae{} from the Counting
\MSO fragment of Presburger \MSO, but extended with regular expressions for 
matching data values:
\[
  \#(*".tex":\{\Document:\{\}\})=1
\]
Alternatively, any formula of Presburger \MSO can be expressed by a 
Presburger tree automaton \cite{SeidlSchwentickMuscholl03,BonevaTalbot05},
if extended with regular expressions for matching data values.

The existing notions of tree automata for unordered trees are well-investigated 
concerning expressiveness (either Counting or Presburger \MSO: \CMSO or \PMSO)  \cite{SeidlSchwentickMuscholl03,BonevaTalbot05},
and have the advantage that membership can be tested in \PTime. When it comes to static analysis
problems such as satisfiability, inclusion, or equivalence checking, they all lack a proper notion 
of determinism, which makes it difficult to distinguish subclasses of automata for which these
problems can be solved efficiently. An exception is the class of feature automata \cite{NiehrenPodelski93}
which have the same expressiveness as \CMSO, but these have the disadvantage that they grow exponentially 
in size for testing simple patterns such as
\(
  \{\Textm{a_1}:\set{},\  \Textm{a_2}:\set{},\ldots,\Textm{a_n}:\set{}\}
\).
The problem is that feature automata must be able to read the \(n\) different edge labels in 
all possible orders.

In this paper, we introduce a general framework for defining
classes of bottom-up automata for unranked unordered trees, 
that abstracts from the way in which properties of horizontal 
languages are specified. The problem is to 
find a good notion of “horizontal determinism”, since there exists no
order on the children of a node. Rather than using Presburger formul\ae{}
for describing labels of the outgoing edges of a node we shall use 
for this purpose finite automata that rewrite the labels of outgoing
edges of a given node in an arbitrary order. Unfortunately, 
membership testing becomes \NP-hard, since all orders must
be inspected in the worst case. A first notion of horizontal 
determinism can then be defined by a restriction to confluent horiontal 
rewriting, so that the order of rewriting becomes irrelevant. For instance, 
one can test the above arity in the order \(\Textm{a_1},\dots,\Textm{a_n}\)
or else in the inverse order (but not necessarily in all orders
in contrast to feature automata).
Our first positive result is that the restriction to confluent rewriting leads to 
polynomial-time membership, emptiness, and universality, as one might 
have hoped. However, the emptiness of binary intersections as well as inclusion
still suffers from coNP-completeness, which might appear  a little surprising, so 
confluence alone is not sufficient for efficiency. 

A second notion of horizontal determinism can be obtained by imposing a fixed
order on the horizontal evaluation, globally for all automata in the class. 
Depending on the choice of the order, we obtain different classes of 
automata but all of them have the same expressiveness, which is that of \CMSO.
We show that this leads to polynomial time membership, emptiness,
universality, emptiness of binary intersections, equivalence, and 
inclusion problems.

\paragraph{Outline}
In Section \ref{sec:prelim}, we recall the notions of automata for
ranked ordered trees, unordered data trees, and Presburger formul\ae{}.
In Section \ref{sec:automata}, we introduce a general framework
for defining classes of bottom-up automata for unordered trees. 
In Section \ref{sec:pres}, we instantiate our framework
for introducing alternating Presburger tree automata, In
Section \ref{sec:horizontal}, we discuss alternating tree automata with
horizontal rewriting, and in Section \ref{sec:confluence} the                                          
restriction to confluent rewriting. Automata for fixed-order
rewriting are introduced in Section \ref{sec:fixed order}.

\section{Preliminaries}
\label{sec:prelim}
\subsection{Automata on Ranked Ordered Trees}

We recall here the classical model of tree automata on ranked trees (\cf \cite{Tata07} for an introduction). 
A ranked signature is a set \(\Sigma\) of function symbols, each of
which has a fixed arity \(\ar(f)\in\N\). A ranked tree is
a term \(t\) with the abstract syntax \(t\affectEQQ f(t_1,\ldots,t_n)\)
where \(n=\ar(f)\).

\mk

\begin{defn} An \emph{alternating (bottom-up) tree automaton for ranked trees}
is a tuple \(B=(\Sigma,\Q,\fin\Q,\Rules)\) where
\(\Sigma\) is a finite ranked signature, \(\Q\) a finite set of states, \(\fin\Q\subseteq \Q\)
the set of final states, and \(\Rules\) a finite set of rules of the form
\(\psi \to q\) where \(\psi\) is a formula with the abstract syntax 
\(\psi \affectEQQ f(q_1,\ldots,q_n) \mid \psi \wedge \psi' \mid \neg \psi\)
for \(f\in\Sigma\) of arity \(n\in \N\) and \(q,q_1,\ldots, q_n\in Q\).
A \emph{nondeterministic  tree automaton for ranked ordered trees}
is an alternating tree automaton, whose rules are of the 
form \(f(q_1,\ldots,q_n)\to q\). A deterministic (bottom-up) 
tree automaton is a nondeterministic tree automaton in which
no two rules share the same left-hand side.
\end{defn}

The evaluator of a nondeterministic tree automaton is defined by
\(\eval B{f(t_1,\ldots,t_n)} = \{ q \mid  q_1\in \eval B {t_1}, \ldots,\) \(q_n\in \eval B {t_n},\ (f(q_1,\ldots,q_n) \to q)\in\Rules\}\). The language defined by $B$ is 
$\{t \mid \eval B t \cap \fin\Q\not=\emptyset\}$. For instance, consider
the set of Boolean formulas $t \affectEQQ \Ltrue \mid \Lfalse \mid \Land(t,t)$.
The set of all valid Boolean formulas can the be defined by 
the deterministic tree automaton with state set $\Q=\{0,1\}$, 
final states $\fin\Q=\{1\}$ and rules 
$\Ltrue \to 1$, $\Lfalse\to 0$, and
$\Land(1,1)\to 1$.

In order to define an evaluator for more general alternating tree automata,
we define the satisfaction relation \(f(Q_1,\ldots,Q_n)\models \psi\)
by \(f(Q_1,\ldots,Q_n)\models g(q_1,\ldots,q_m)\) iff \(m=n\), \(g=f\), 
and \(q_i\in Q_i\) for all \(1\le i\le n\), extended to negations
and conjunctions as usual, i.e., \(f(Q_1,\ldots,Q_n)\models \psi\wedge \psi'\)
iff $f(Q_1,\ldots,Q_n)\models \psi$ and \(f(Q_1,\ldots,Q_n)\models \psi'\),
and \(f(Q_1,\ldots,Q_n)\models \neg\psi\) iff not \(f(Q_1,\ldots,Q_n)\models \psi\).
As before, the language defined by $B$ is 
$\{t \mid \eval B t \cap \fin\Q\not=\emptyset\}$.

It is well known that alternating, nondeterministic, and deterministic tree automata 
can define the same classes of languages of ranked trees, which are those
definable in \MSO.

\subsection{Unordered Unranked Data Trees}

An \emph{alphabet} is a finite set \(\Alphabet\). A \emph{data value} 
over \(\Alphabet\) is a string in \(\Alphabet^*\). We write \(d_1d_2\) for the concatenation
of strings \(d_1,d_2\in\Alphabet^*\). 

Let \(\Nat\) be the set of natural numbers including \(0\). 
A \emph{multiset} over a finite set \(D\) is a function \(M:D\to\N\). 
The set of multisets over \(D\) is written \(\MS(D)\).
As usual, we write \(\b{d_1,....,d_n}\) for the multiset in which
each element of \(d\in D\) has the same multiplicity as the number
of occurrences of \(d\) within the brackets. Given a second set \(X\), 
we use record notation for multisets over pairs in \(D\times X\),
i.e., we  write \(\b{d_1:x_1,\ldots,d_n:x_n}\)
instead of \(\b{(d_1,x_1),\ldots,(d_n,x_n)}\) for any
\(d_i\in D\) and \(x_i\in X\) -- and sometimes use the same notations
for isolated pairs \(d:X\).

We define the set \(\Trees\) of 
\emph{unordered, edge--labelled data trees} 
(or simply \emph{trees} in this paper) 
over data alphabet \(\Alphabet\) 
inductively as the least set that contains all 
multisets \(\b{d_1:t_1,\ldots, d_n:t_n}\) such
that \(n\> 0\), \(d_1,\ldots,d_n\in\Alphabet^*\) and \(t_1,\ldots,t_n\in\Trees\).
Given a tree \(t=\b{d_1:t_1,\ldots, d_n:t_n}\), the 
multiset \(\b{\enum d1n}\) is called the \emph{arity} of \(t\).

\begin{figure}[t]
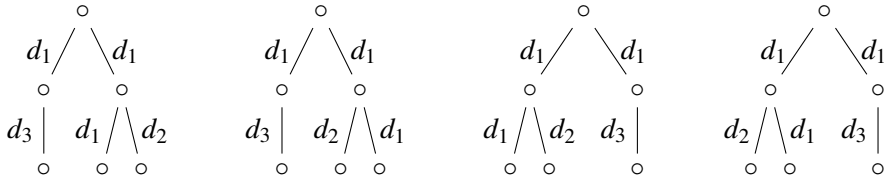

\[
\tzt [t,ps]{\circ}{
   \edge node [left] {d_1}; [.{\circ}
     \edge node [left] {d_3};{\circ} ]
   \edge node [right] {d_1}; [.{\circ}
     \edge node [left] {d_1};{\circ}
     \edge node [right] {d_2};{\circ} ]}
\qquad
 \tzt [t,ps]{\circ}{
   \edge node [left] {d_1}; [.{\circ}
     \edge node [left] {d_3};{\circ} ]
   \edge node [right] {d_1}; [.{\circ}
     \edge node [left] {d_2};{\circ}
     \edge node [right] {d_1};{\circ} ]}
\qquad
\tzt [t,ps]{\circ}{
   \edge node [left] {d_1}; [.{\circ}
     \edge node [left] {d_1};{\circ}
     \edge node [right] {d_2};{\circ} ]
   \edge node [right] {d_1}; [.{\circ}
     \edge node [left] {d_3};{\circ} ]}
\qquad
\tzt [t,ps]{\circ}{
   \edge node [left] {d_1}; [.{\circ}
     \edge node [left] {d_2};{\circ}
     \edge node [right] {d_1};{\circ} ]
   \edge node [right] {d_1}; [.{\circ}
     \edge node [left] {d_3};{\circ} ]}
\]
\caption{\label{tree-graphs} Drawings of  \(\b{
  {d_1}: \b{{d_3}: \b{}},
  {d_1}: \b{
    {d_1}:\b{}, {d_2}:\b{}
  }
}\) with different edge orders.}
\end{figure}

We employ the usual graphic representation where
\(\b{
  {d_1}: \b{{d_3}: \b{}},
  {d_1}: \b{
    {d_1}:\b{}, {d_2}:\b{}
  }
}\)
is drawn as one of the many graphs in Figure \ref{tree-graphs}
that differ only in the order in which the outgoing edges 
of nodes are drawn. Note that each node in a
tree has a finite, but unbounded, number of sons.



We will use regular expressions as pattern for matching data values.
A regular expression  \pat\ has the following abstract syntax
where \(d\in\Alphabet^*\):
\[
  \pat \affectEQQ \Text{d} \mid \pat\pat  \mid \pat+\pat \mid \pat^* 
  \mstop
\]
The set of regular expressions \(\pat\) is denoted by \(\PatREG\). 
The semantics of a pattern \(\pat\in\PatREG\) is a set 
of data values \(\sem{\pat}\subseteq \Alphabet^*\) defined 
in the classical manner \cite{hopcroft2001introduction}.
As syntactic sugar, we let
\(* \equiv (a_1+\dots+ a_m)^*\), where \(\Alphabet=\{a_1,\dots,a_m\}\).

\subsection{Descriptor Classes and Presburger Formul\ae}\label{sec:presbintro}

\begin{defn}[Descriptor Class]
A \emph{descriptor class} for a set \(\Models\) of models is a tuple \(\tuple{\Desc,\models,\size{\cdot},\cost}\) where
\(\Desc\) is a set of \emph{descriptors}, \(\models\)
a subset of \(\Models \times \Desc\), \(\size{\desc}\in\N\) the \emph{size} of
a descriptor \(\desc\in\Desc\), and \(\cost\in\N\) the \emph{cost} of the class.
\end{defn}

As a first example, 
any subset \(\Pat\incl\PatREG\) of regular expressions over our data alphabet \(\Alphabet\)
can be seen as a descriptor class selecting words in \(\Alphabet^*\):
satisfaction is defined as \(d\models\pat\) iff \(d\in\sem{\pat}\),
the size \(\size\pat\) of a pattern \pat\ is the number of its symbols, 
and the cost of the class is \(\cost=0\).

\label{sec:presburger-atoms}


We recall the definition of propositional 
Presburger formul\ae{}, which serve to specify properties 
of multisets. The logic is parametrized by a set \(X\) over which the
multisets are constructed and a descriptor class \(\tuple{\Filters,\models,\size{\cdot},\cost}\) 
providing descriptors for elements of \(X\) that we shall call \emph{filters}.

Presburger formul\ae{} \(\presb\) are built from filters as follows. 
One first constructs counting expressions \(\nu\), which are
either constants \(n\in\Nat\), sums \(\nu+\nu'\), or counters for filters
\(\#\f\) where \(\f\in\Filters\). A counter \(\#\f\) sums up the multiplicities 
of all elements of the multiset satisfying \(\f\):
\[
\begin{array}{rcl}
\nu  &\affectEQQ& n \hmid \# \f \hmid \nu+\nu \\
\presb &\affectEQQ& \nu~\le~\nu' \mid \nu\eqmod m \nu' \mid
\presb\wedge\presb'\mid \neg \presb
\end{array}
\]
An atomic Presburger formula \( \nu~\le~\nu'\)  or \( \nu~\eqmod m~\nu'\) compares the
values of two counting expressions. General Presburger formul\ae{} are
constructed from atomic Presburger formul\ae{} and the usual Boolean
operators from propositional logic.  
Given a multiset \(M\) over \(X\), the semantics \(\sem{\nu}^M\in\Nat\)
is defined as usual:
\[
\begin{array}{lll}
\sem{n}^M = n,\qquad\qquad
&
 \sem{\nu+\nu'}^M= \sem{\nu}^M+\sem{\nu'}^M,\qquad\qquad
&
\sem{\#\f}^M= \sum_{x\models \f} M(x) 
\end{array}
\]
We say that \(M\) satisfies the atomic formula \(\nu~\le~\nu'\) if
\(\sem{\nu}^M~\le~\sem{\nu'}^M\) and similarly \(M\) satisfies 
\(\nu\eqmod m \nu'\) if \(\sem{\nu}^M=\sem{\nu'}^M\mod m\).
This extends to Boolean combinations in the usual way.
In this case we write \(M\models \f\).

Presburger formul\ae{} with filters in \(\Filters\) define a 
descriptor class for multisets over \(X\). The size of a Presburger formula is the sum of the number of its
symbols, excepting filters, plus the sizes \(\size{\f}\) of all occurrences of filters
\(\f\) in the formula. The cost of the class of Presburger formul\ae{} is the
cost of its class of filters.

\section{Automata for Unordered Trees}
\label{sec:automata}

We start with abstract classes of bottom-up automata for unordered
unranked trees, which generalize on alternating tree automata as well as on
nondeterministic tree automata. This will enable us to introduce
alternating Presburger automata (in Section
\ref{sec:pres}) and alternating tree automata with horizontal rewriting
(in Section \ref{sec:horizontal}) as concrete instances.

\subsection{Automata for Unordered Trees}

We fix \QQ, a countable set of \emph{properties}.
We develop a parametrized framework of automata, in which one can freely choose
a descriptor class for matching arities that are 
decorated with sets of properties, which will also be sets of our automata's states.
\medskip

\begin{defn}
A \emph{horizontal descriptor class} \H\ 
is a descriptor class for multisets over \(\Alphabet^*\times \powerset\QQ\).
\end{defn}
The \emph{support} $\Supp (h)$ of a horizontal descriptor $h\in\H$ 
is the set of all properties that $h$ actually deals with;
it is defined as the least subset $\Q$ of $\QQ$
such that
for any \(i=1..n\), \(d_i \in \Data\) and \(Q_i,Q_i'\incl \QQ\) such that 
\(Q_i'\inter \Q = Q_i\inter \Q\),
\(\b{d_1:Q_1,\ldots,d_n:Q_n}\models h \iff
\b{d_1:Q'_1,\ldots,d_n:Q'_n}\models h\).\mk
%



\begin{defn}[\AUTs]
An \emph{alternating bottom-up automaton for unordered unranked data trees%
} (\AUT) is a tuple \(A=\tupleAUT\) where
\(\Q\incl \QQ\) is the finite set of \emph{(vertical) states}, 
\(\fin \Q \subseteq \Q\) the subset of final states,  
\(\Horizontals\) is a horizontal descriptor class, 
and \(\Rules\incl \Horizontals \times \Q
\), such that for all $(h,q)\in \Rules$, $\Supp(h)\incl \Q$,
is the finite set of \emph{(vertical) transition rules}.
\end{defn}

We shall write \(\horizontal\to q\) if \((\horizontal,q)\in\Rules\). 
Any automaton \(A\) evaluates any tree with data alphabet \(\Alphabet\) to a set of
states. This set is defined by induction on the structure
of trees such that for all \(n\ge 0\), data values 
\(d_1,\ldots, d_n \in\Alphabet^*\) and trees \(t_1,\ldots, t_n\in\Trees\):
\begin{align*}
\eval A{\b{d_1:t_1,\ldots,d_n:t_n}} &= \setst{ q }{  \b{d_1:\eval A {t_1},\ldots, d_n:\eval A {t_n}}\models\horizontal,\ \horizontal \to q}
\mstop
\end{align*}
Alternation requires to consider all states 
assigned to subtrees when applying a transition rule, and not
only one of them nondeterministically.
The \emph{language accepted by \(A\)} is defined as
\(\L(A) = \setst{t\in \Trees}{ \eval At \inter \fin \Q \ne \void}\).
The size \(\size A\) is the sum of the number of states \(\card \Q\), the size
\(\sum_{\horizontal\to q}1+\size\horizontal\),
and the cost of the descriptor class \(\Horizontals\).\mk


%

\begin{defn}
The \emph{class} \AUTFun{\H} is the set of all \AUT\ 
whose horizontal descriptor class is \H.
\end{defn}

As a first example, we consider the horizontal descriptor
class \(\HArity{}\), which tests arity constraints. 
An arity
constraint has the form
\(
   \b{"d_1":q_1,\ldots,"d_n":q_n} \mcom
\) 
where \(n\ge 0\), \(q_1,\ldots,q_n\in \QQ\) and \(d_1,\ldots,d_n\in\Alphabet^*\).
It is satisfied by all multisets of the form \(M+\b{"d_1":Q_1,\ldots,"d_n":Q_n}\)
such that \(q_i\in Q_i\) and \((d_i,Q)\not\in M\) for all \(1\le i \le n\) and any \(Q\). 
The size of an arity constraint is the number 
of its symbols. The  cost of any descriptor class  \(\HArity{\Q}\) is \(0\).

As a second example, we consider the richer class of horizontal 
descriptors \(\HArityAnd{}\) which, besides arity constraints, supports Boolean 
operators, \ie the formul\ae{} \(\psi\) of \(\HArityAnd{\Q}\) are given by the 
following abstract syntax, where all \(q_i \in Q\) and \(d_i\in \Alphabet^*\):
\[
\psi \affectEQQ \b{"d_1":q_1,\ldots,"d_n":q_n} \mid \psi \wedge \psi' \mid \neg\psi
\mstop
\] 
The automata from the classes \(\AUTFun{\HArity{}}\)
and \(\AUTFun{\HArityAnd{}}\) show how easy it is to translate
the notions of ranked automata into the unordered framework.
More precisely:

\ignore{
See
Prop. \ref{alternatingnondetexpressivepower} for a precise comparison.


The lack of alternation of \(\AUTFun{\HArity{}}\) is due to
the omission of conjunction in the formul\ae{} of \(\HArity{\Q}\).
More precisely, it is due to the fact that for any \(\Q\) 
all formul\ae{} \(\psi\) of \(\HArity{\Q}\) 
satisfy the following \emph{singleton-projection property}:
\[
\b{d_1:Q_1,\ldots,d_n:Q_n}\models\psi \xxx{iff} \exists q_1\in Q_1,\ldots, q_n\in Q_n : \b{d_1:\{q_1\},\ldots,d_n:\{q_n\}}\models\psi
\]
In contrast, this property fails for formul\ae{} with conjunction such as \(q\wedge q'\) 
but also for Presburger formul\ae{} without conjunction such as \(\#q+\#q'\ge 2\).
Therefore, automata \(\AUTFun{\HArity{}}\) can be evaluated in
in a nondeterministic manner on any tree, by guessing states 
\(q_1\in Q_1,\ldots, q_n\in Q_n\) and testing whether \(\b{d:\{q_1\},\ldots,d_n:\{q_n\}}\models\psi\).
This justifies the following semantic definition of alternation-freeness:

\medskip
\begin{defn}[Alternation-freeness]\label{def:af}
We call an \AUT \(\tupleAUT\)
\emph{alternation-free} or equivalently \emph{nondeterministic} if the singleton-projection property is satisfied by all its
horizontal descriptors of \(\Horizontals\).
\end{defn}

It should be noted that the traditional syntactic notions of nondeterministic automata 
correspond to particular \AUT classes such as \(\AUTFun{\HArity{}}\). As argued 
above, all automata of such classes are alternation-free in the sense of Definition \ref{def:af}.

%

In order to relate \AUTs to traditional classes of tree automata,
we show that the class \AUTFun{\HArity{}} captures 
nondeterministic tree automata for ranked ordered trees, and
that the more general class \AUTFun{\HArityAnd{}} captures
alternating tree automata for ranked ordered trees. \mk
} 

\mk
\begin{prp}[Encoding Automata for Ranked Ordered Trees]
\label{alternatingnondetexpressivepower}
There exists an encoding \(\sem\cdot\) of ranked ordered trees into
unordered trees, and of alternating
ranked ordered tree automata into 
\(\AUTFun{\HArityAnd{}}\), such that
for any automaton \(B\) on ranked ordered trees we have
\(\sem{\L(B)}=\L(\sem{B})\).
Furthermore, if \(B\) is non-deterministic, then \(\sem{B}\in\AUTFun{\HArity{}}\).
\end{prp}
\ifLongPaper
\begin{proof}
We assume a finite ranked signature \(\Sigma\) of function symbols.
To encode these function symbols we fix the alphabet
\(
   \Alphabet = \{0,\ldots,9,a,\ldots,z\}
\)
and assume that any symbol \(f\in \Sigma\) can be mapped to some data value 
\(\sem{f}\in\Alphabet^*\setminus\{0,\ldots,9\}^*\) in a unique but arbitrary 
manner. In order to encode ranked ordered trees of \(\Sigma\) into unordered 
trees over \(\Alphabet\), we map natural numbers \(n\in\Nat_0\) to their 
usual decimal representation \(\sem{n}\in\{0,\ldots,9\}^*\), \ie \(\sem{51}="51"\)
and lift this mapping to ranked trees as follows:
\[
  \sem{f(t_1,\ldots,t_n)} =  \b{\sem{f}:\b{},"1":\sem{t_1},\ldots,\sem{n}:\sem{t_n}}
\]

We can encode an alternating tree automaton \(B\) for ranked ordered trees
into an automaton \(\sem{B}\) in \AUTFun{\HArityAnd{}} as 
follows. Automaton \(\sem{B}\) has the state set \(\Q\uplus\{q_\LEAF\}\). It has
one transition rule for leaves \( \b{} \to q_\LEAF\), plus one transition rule
for each rule of \(B\) obtained by encoding the formula
on the left hand side of the rule as follows:
\[
\begin{array}{l}
\sem{f(q_1,\ldots,q_n)} =\b{\sem{f}:q_\LEAF,"1":q_1,\ldots,\sem{n}:q_n},\qquad
\sem{\psi\wedge\psi'} =\sem{\psi}\wedge\sem{\psi'},\qquad 
\sem{\neg\psi}  =\neg\sem\psi
\end{array}
\]
Clearly, the size of a rule is preserved by the translation up to a constant factor.
Furthermore, the language of an automaton is preserved modulo encoding of ordered ranked into unordered
trees. It is also obvious that logical operators appear in the encoding if and only if
they do in the original: the encoding preserves alternation.
\end{proof}
\fi

\subsection{Complexity}

\begin{prp}[Membership]\label{general-membership}
Let $C$ be a class of automata such that for all
$\tupleAUT\in C$ and \(h\in\H\),
whether \(\b{d_1:Q_1,\ldots,d_n:Q_n}\models \h\) 
can be decided in time  \(O(\sum_{m=1}^n \size{d_m}\mult \size h)\),
for any \(\enum d1n\in\Alphabet^{*}\),
and finite sets \(\enum Q1n\subseteq \QQ\).
In this case, membership \(t\in \L(A)\) for trees \(t\in \Trees\) and
automata \(A\in C\) can be decided in time \(O(\size t \mult \size A)\).
\end{prp}
\ifLongPaper
\begin{proof}
Let \(k\in\Nat\) such that whether \(\b{d_1:Q_1,\ldots,d_n:Q_n}\models \h\) 
can be decided in time at most \(k\mult \sum_{i=1}^n \size{d_i}\mult \size h\). It is sufficient
to show that we can compute \(\sem{t}_A\) in time at most \(k\mult \size t \mult \size A\).
The proof is by induction on the structure of \(t\). 
If \(t=\b{d_1:t_n,\ldots,d_n:t_n}\)
then \(\eval A{t} = \setst{ q }{  \b{d_1:\eval A {t_1},\ldots, d_n:\eval A {t_n}}\models\horizontal,\ \horizontal \to q}\).
We compute the sets \(Q_i = \sem{t_i}_A\) in time \(k \mult \size{t_i} \mult \size A\)
for all \(1\le i\le n\). We then test \(\b{d_1:Q_1,\ldots,q_n:Q_n}\models h\)
in time \(k \mult \sum_{i=1}^n \size{d_i}\mult \size h\)  for all rules \(h\to q\),
and thus in time at most \(k \mult \sum_{i=1}^n \size{d_i}\mult \size A\).
The overall time is at most \(k \mult \size{t}\mult\size A\).
\end{proof}
\fi

A descriptor class \(\H\) is closed by the boolean operation \(\circledast\) if
for every \(h,h'\in\H\), there is \(h\circledast h'\in\H\) such that
\(M\models h\circledast h'\) iff \((M\models h) \circledast (M\models h')\).\mk

{\nc\all[1]{\x{all}_{#1}} 
\begin{prp}[Emptiness]\label{general-emptiness}
Let \H\ be such that
\point1 for any \(h\in\H\), 
whether \(\exists M :M\models \h\) is decidable in time  \(O(g(\size \h))\),
and \point2
\H\ is  closed by all boolean operations in linear time, and
\point3
for any  \(\Q\incl \QQ\) and \(S\incl \powerset\Q\), there exists \(\all S \in \H\)
of size \(O(2^{\card\Q})\)  that is satisfied exactly by all multisets over \(\Alphabet^*\times S\).
In this case, whether \(\L(A)=\void\) can be decided in time 
\(O(
2^{2\cdot{\card \Q}} \cdot g(2^{\card \Q}  + \size A)
)\) for all automata \(A\in\AUTFun\H\) of states \Q.
\end{prp}}%

\ifLongPaper
\begin{proof}
\nc\all[1]{\x{all}_{#1}} 
\nc\reach[1]{\x{reachable}(S,#1)}%
We use a variation on the usual reachability algorithm for 
bottom-up tree automata; because of the alternation, we need to build iteratively a set 
\(S \incl \powerset\Q\) of reachable annotation sets.
We initialize the algorithm with \(S:=\emptyset\) and 
we iterate as follows. At each step, we add to \(S\)
the new annotation sets that have become reachable thanks to \(S\).
This proceeds much as for the initialisation phase,
with horizontal descriptors instead of patterns,
but this time we need to restrict the satisfiability checking
to using only sets in \(S\). 
To do that we use the descriptor \(\all S\)
provided in our hypotheses, such that
\[
\b{d_1 : Q_1, \dots , d_n: Q_n} \models \all S
\xx{iff}
\forall 1 \< i \< n,\ Q_i \in S
\mstop
\]
Whether and how this is coded depends on the concrete class of descriptors;
as an example, in the Presburger logic defined in the next section,
we could write
\[
\all S\ = \quad
\#\px[\Bigg]{
\lOr_{Q\notin S}
\pbx[\bigg]{
\lAnd_{q \in Q }q \land \lAnd_{q \notin Q }\neg q
}
} = 0
\mcom
\]
where \(q\) is a primitive that tests the presence of a state in an annotation set.
Note that the size of that is at most exponential, as required in the hypotheses.

Whatever the implementation, using \(\all S\) as a primitive, 
we define the predicate \(\reach Q\), true iff
\(Q\) is reachable in one bottom-up transition, starting from
a set \(S\) of reachable annotation sets.
That is the case if there exists an annotated arity \(M\),
whose annotation sets are all in \(S\),
such that all states in \(Q\), and exactly those,
are right-hand sides to at least one non-leaf rule
whose horizontal descriptor is satisfied by \(M\). 
Thus the predicate is expressed by the satisfiability of a descriptor:
\[
\reach Q\ = \quad
\px[\Bigg]{
\all S
\llland
\lAnd_{q\in Q} \lOr_{\h\to q} \h 
\llland
\lAnd_{\substack{q\notin Q\\\h\to q}} \neg\h}
\mstop
\]
A step of the reachability algorithm
tests all possible sets of states, and adds them
to \(S\) if they are reachable:
\[
S \affectEQ S \union
\setst{
  Q \incl \Q
}
{ \exists M.\ 
 M\models \reach Q
}\mstop
\]
In a single reachability step, there are therefore \(2^{\size \Q}\) satisfiability tests to make,
each one doable in \(O(g(2^{\size \Q}  + \size A))\).
This operation is repeated until no new \(Q\) can be made accessible
-- or until a final state appears in one of the reachable annotation sets --
thus it cannot be executed more than \(2^{\size \Q}\)
times in total (any reachable \(Q\) stays reachable). 
Therefore, in total, we have
\(O(
2^{\size \Q}\cdot 2^{\size \Q} \cdot g(2^{\size \Q}  + \size A)
)\)
for all the iteration steps.
\end{proof}
\else
\begin{proof}
We perform a vertical reachability algorithm on sets of 
simultaneously reachable states. Each step involves testing
all state subsets, and each test is exponential.
There are at most an exponential number of steps,
as each reachable subset remains reachable throughout.
\end{proof}
\fi

Note that under the conditions of that proposition,
the boolean closure properties for the automata, and the decidability
of universality, disjointness, equivalence, and inclusion
follow naturally, some technicalities notwithstanding.
In a nutshell, one must be careful any time two descriptors acting on different sets
of states must interact or relate to one another.
We also need more sophisticated notions of boolean closure, for families of descriptor classes.
Those details are more tedious than difficult, and are left out of this paper.

%
%

\subsection{Vertical Determinism}

We next introduce the notion of vertical determinism,
which is the ``standard'' view of determinism in bottom-up automata:
that is to say, trees are evaluated in at most one state.
Like in the ranked case, vertical determinism is necessary in order to obtain good complexities
for static analysis problems. It is not sufficient, however: all the classes which we consider in the next
sections define filters (see Sec. \pref{sec:presbintro}) that can manipulate regular patterns and properties,
by conjunction, disjunction or negation. If all regular patterns were allowed, testing satisfiability
of such filters would be \PSPACE-hard (by emptiness of intersection of regular languages). And if the automata were alternating, 
\NP-hardness would be hard to avoid, as sets of properties are tested, which can encode variable assignments,
for instance (SAT problem).
To get reasonable complexity and reasonable expressive power, one must therefore
combine vertical determinism and restrictions on the patterns one can test.

Fortunately, provided that \H\
satisfies the boolean closure properties,
any \AUTFun\H can be transformed into an equivalent,
vertically deterministic \AUTFun\H. 
\mk

\begin{defn}[Vertical Determinism]
An \AUT\ \(A\) is \emph{vertically deterministic} if
\(\max_{t\in\Trees} (\card{\eval At}) = 1\).
\end{defn}\mk


\begin{prp}[Vertical Determinisation]\label{general-determinisation}
For any \(A\in \AUTFun\H\), an equivalent vertically deterministic \(B\in\AUTFun\H\) 
can be constructed in time \(O(2^{2\size A})\),
provided that \H\ is closed in linear time 
under conjunction and negation.
\end{prp}
\ifLongPaper
\begin{proof}
Let \(A=\tupleAUTqJ\); then
\(A'=\tuple{\Alphabet,\powerset \Q,\fin \Q',\HFun{\powerset\Q},\Rules'\union \Rules''}\),
where 
\begin{align*}
\fin \Q' &= \setst{Q\incl \Q}{Q\inter \fin\Q \ne \void} \ttand
\\
\Rules'&= 
\setst{
\p{
\quad\so{\lAnd_{\pat\to q \in X}} \pat
\llland 
\so{\lAnd_{\pat\to q \notin X}} \neg \pat
}
\to
\setst{q}{\pat\to q \in X}
}{X\incl \Rules\inter (\RulesP)} \mcom
\\
\Rules''&= 
\setst{
\p{
\quad\so{\lAnd_{\h\to q \in X}} \bar \h
\llland 
\so{\lAnd_{\h\to q \notin X}} \neg \bar \h
}
\to
\setst{q}{\h\to q \in X}
}{X\incl \Rules\inter (\HFun{\Q}\times \Q)}
\mstop
\end{align*}
Note that by construction no two rules of \(A'\) may apply on the same term, and that
for all \(t\), \(\eval {A'} t = \sset{\eval At}\), which makes \(A'\) deterministic 
and complete. As for the complexity, we build of the order of \(2^{\size A}\) rules,
each containing \(O(2^{\size A})\) descriptors, each of size at most \(f(\size A)\).
\end{proof}
\fi



\section{Alternating Presburger Tree Automata}
\label{sec:pres}

We introduce \emph{alternating Presburger automata} for unordered unranked
data trees (\AUTPs), by instantiating the horizontal descriptors of \AUT's
by propositional Presburger formul\ae{}, and present expressiveness
and complexity results.


We now define the descriptor class \(\HFunP{}\) 
of propositional Presbuger formul\ae{}. 
We first fix a descriptor class \(\Pat\), subclass of \PatREG, for words in \(\Alphabet^*\).
We then define the filters \(\f\) with the 
following syntax, where \(\pat\) is a descriptor of \(\Pat\) and \(q\in \QQ\):
\[
  \f \affectEQQ \pat \mid q \mid \f \wedge \f \mid \neg \f
  \mstop
\]
The semantics is defined as follows, for
\((d,Q)\in \Alphabet^*\times \powerset\QQ\):
\((d,Q)\models q\) iff \(q\in Q\),
\((d,Q)\models\pat\) iff \(d\models \pat\). 
The inductive cases are as usual. The size of a filter \(\size \f\) 
is the number of its symbols plus \(\size\pat\)
for all occurences of \(\pat\). The cost of the filter
class is the cost of the pattern class, \ie \(0\).

\medskip
\begin{defn}[\AUTP: Alternating Presburger Tree Automata] 
The class \AUTP of \emph{alternating bottom-up Presburger automaton for unordered unranked
trees} is defined as \AUTFun{\HFunP{}}.
\end{defn}

Alternating Presburger automata take into account the fact that a tree
may be recognized in several states. For instance, \(
\b{d_1:\{q_1,q_2\}, d_2:\{q_2,q_3\}} \models \#q_1+\#q_2=3\). This
allows in general to obtain more concise automata than in case of the
Presburger tree automata of
\cite{SeidlSchwentickMuscholl03,BonevaTalbot05} which are
non-deterministic, which is to say that acceptance is based on the notion of
a run that assigns a single state to each tree, even when this is done
in a non-deterministic way.  

As a consequence, \AUTP\ do not directly capture all Presburger
tree automata from \cite{BonevaTalbot05}, but only those that are
alternation-free. From the viewpoint of expressiveness, this is good
enough, since vertically deterministic \AUTPs capture \PMSO already,
as we shall see in Theorem \ref{theo:pmso}.

\medskip
\begin{expl}
\label{sec:latex}
Let us illustrate the above by showing an \AUTP\ checking some basic
cleanness criteria for a \LaTeX\ document directory.
We require that the files produced by compilation,
\ie all files whose name matches \texttt{*.dvi}, \texttt{*.pdf}, \texttt{*.aux},
are absent from the directory.
Furthermore, all \texttt{*.tex} must be simple files, and
exactly one among them must be a valid \TeX{} main document.
There is no restriction on the subdirectories.
We note \(\xlbl \pi {main}=\Documentclass\ast\), \(\xlbl \pi {cmp}=\ast\Text{.dvi}+\ast\Text{.pdf}+\ast\Text{.aux}\).
We have the following rules:

\begin{tabular}{r r l}
\(\#(\ast)=0\) & \(\to\) & \(\ql{leaf}\)\\
\(\#(\xlbl \pi {main} \land \ql{leaf})=1\land \#(\ast)=1\) & \(\to\) & \(\ql{main}\)\\
\(\#(\ql{leaf})=1\land \#(\ast)=1\) & \(\to\) & \(\ql{file}\)\\
\(\#(\ast\Text{.tex}\land \ql{main})=1\land \#(\ast\Text{.tex}\land \lnot\ql{file})=0\land \#(\xlbl \pi {cmp})=0\) & \(\to\) & \(\ql{ok}\)\\
\end{tabular}

State \(\ql{leaf}\) is assigned to leaf nodes, and
state \(\ql{file}\) to all nodes representing files, 
i.e. nodes with exactly one outgoing edge, whose data value 
is the file's content and whose target is a leaf node.
State \(\ql{main}\) is assigned to all nodes with
one outgoing edge labeled by the content of a main
\LaTeX\xspace file, i.e., a string  matching \(\Documentclass\).
State \(\ql{ok}\) accepts only clean \LaTeX\ repositories.

Note that the properties tested by \(\ql{main}\) and \(\ql{file}\) are not mutually exclusive.
To make this automaton alternation-free, we would have to force \(\ql{file}\) to specifically test that its only data value doesn't match \(\Documentclass\).
Here, alternation facilitates specification.
\end{expl}

\label{sec:pmso}
We now present a logical characterization of the expressiveness of \AUTP's,
showing that they capture \PMSO.
We assume a possibly infinite set \(\mathcal Q\) of set variables ranged over by \(q\). Let \(\presb\) range over propositional Presburger formul\ae{} with predicates in \(\mathcal Q\), which describe multisets over \(\Alphabet^* \times 2^{\mathcal Q}\). The formul\ae{} \(\alpha\) and \(\beta\) of \PMSO are then defined by:
\[
\begin{array}{ll}
\mbox{node sets}    &  \alpha \affectEQQ \presb \mid q \mid \{root\} \\
\mbox{truth values} &  \beta \affectEQQ \alpha\subseteq \alpha'  \mid \beta\wedge \beta' \mid \neg \beta\mid  \forall q. \beta 
\end{array}
\]
A formula with free variables in \(\Q\subseteq \mathcal Q\) can then be interpreted over an
unordered tree \(t\) and a assignement \(\sigma\) of \(\Q\) to set of nodes of \(t\). This defines
a satisfaction relation \(t,\sigma\models \beta\).

\medskip

\begin{thm}[\AUTP Expressiveness]
\label{theo:pmso}
A language of unordered data trees is definable by an \AUTP
if and only if it is definable by a
closed \PMSO formula with the appropriate alphabet.
\end{thm}

\begin{proof}[Proof sketch]
Let \(A\) be an \AUTP. We translate rules \(\presb \to q\) by \(\presb\subseteq q\). 
We then impose that all sets \(q\) 
are minimal while satisfying all the \PMSO formul\ae{} for the rules. Finally,
we impose that \(\vee_{q\in \fin\Q} \{root\}\subseteq q\), and quantify existentially
over all predicates.
Conversely, we can show that the non-alternating Presburger tree automata from 
Boneva and Talbot \cite{BonevaTalbot05} can be expressed by \AUTP's. Their definition is close to that
of \AUTP's except that the semantics is defined in a non-deterministic manner,
and not alternating. Nevertheless, our formalism subsumes in a natural manner
the subclass of their Presburger tree automata that are horizontally deterministic
Since they capture \PMSO \cite{BonevaTalbot05}, \AUTP's do too (except that string 
patterns are not supported by their version of \PMSO and Presburger automata, but they
can be eliminated in a preprocessing step). Alternatively, 
a direct proof of this result can be obtained as usual when relating \MSO to 
standard tree automata over ranked trees \cite{ThatcherWright68,Tata07}.
\end{proof}


\begin{prp}[\AUTP\ Complexity]\label{autp-complexity}
Given vertically deterministic \AUTP\ \(A,B\) and a tree \(t\in \Trees\), 
deciding whether \(t\in \L(A)\) is \PTIME,
\(\L(A)=\void\) is \PSPACE-hard,
\(\L(A)=\Trees\) is \PSPACE-hard,
\(\L(A)\inter \L(B) = \void\) is \PSPACE-hard.
\end{prp}

\begin{proof}
These complexity results follow from known results on Presburger logic 
\cite{DBLP:conf/icalp/SeidlSMH04,DBLP:conf/birthday/SeidlSM08}.
\ifLongPaper

\textbf{Membership.}
Satisfaction of filters in \HFunP{\Q} is testable in polynomial time,
because satisfaction of a regular expression and membership of
a state in a state set are both obviously \PTIME.
Membership for Presburger formul\ae\
with \PTIME-testable atoms is \PTIME\ itself.
Thus the result follows from Prop. \pref{general-membership}.

\textbf{Emptiness.}
Satisfiability for Presburger on finite alphabets is already
NP-complete \cite{papadimitriou1981complexity},
and that problem can be reduced to emptiness for \AUTP,
by considering languages of trees of height one,
encoding the language horizontally.

\textbf{Universality}
The Universality problem for Presburger constraints
over finite alphabets is already \PSPACE-hard.
\VH{Can't find ref}

\textbf{Disjointness}
Disjointness is at least as hard as emptiness, by taking \(\L(B)=\Trees\).
\fi
\end{proof}


\section{AUTs with Horizontal Rewriting}
\label{sec:horizontal}

We next introduce alternating automata with horizontal rewriting
(\AUTAs) by instantiating \AUTs with ``horizontal'' automata whose 
transitions are guarded by filters. \AUTAs have the same 
expressiveness as \AUTPs but differ in computational 
properties and succinctness. As we shall see in 
the next section, \AUTA make it indeed easier to formulate 
restrictions leading to more efficient static analysis.


Let \(\Filters\) be the set of filters
\(\f\) for words in \(\Alphabet^*\times\powerset \QQ\) from the previous 
section, i.e.,  \(\f \affectEQQ \pat \mid q \mid \f \wedge \f \mid \neg \f\) 
where \(q\in\QQ\) and \(\pi\in \Pat\), where \(\Pat\incl\PatREG\).
\medskip

\begin{defn} A \emph{horizontal automaton} is a triple
\(\tuple{\A,\hStates,\hRules}\) where \(\hStates\) is a finite set of 
\emph{horizontal states} and \(\hRules \subseteq \hStates\times \Filters \times \hStates\) 
is the \emph{horizontal transition relation}.
\end{defn}
We will write \(\hRule{p}{\f}{p'}\) instead of \((p,\f,p') \in \hRules\).
Any horizontal
automaton \(H=\tuple{\A,\hStates,\hRules}\) defines a descriptor class \(\H_H=\tuple{\hStates^2,\models,\size{\cdot},\cost}\) 
for multisets over \(\Alphabet^*\times\powerset \QQ\). Its descriptors are
pairs of horizontal states \((p,p')\in\hStates^2\), where \(p\) serves as an \emph{initial} and 
\(p'\) as a \emph{final} horizontal state of the descriptor. The \emph{horizontal rewriting relation} of \(H\) is the binary 
relation \(\rewh\) on \(\hStates\times \MS(\Alphabet^*\times \powerset \Q)\) 
given by:
\[
(p,M + \b{d:Q}) \rewh (p', M)
\xxx{ if }  
\exists \f: 
\hRule{p}{\f}{p'} \ttand (d,Q) \models \f
\mstop
\]
A multiset \(M\) over \(\Alphabet^*\times \powerset \QQ\) satisfies a descriptor
\((p,p')\) if \(p'\) can be reached from \(p\) while consuming \(M\):
\(
M \models (p,p') \iff  (p,M) \rewh^* (p',\b{})
\mstop
\)
The size of a descriptor \(p,p'\) is \(\size{(p,p')}=2\) while the cost
of the class is the overall size of the horizontal automaton 
\(\cost=\sum_{(p,\f,p')\in\hRules} \size\f\).


\medskip
\begin{defn}[\AUTA]
The class \AUTA of \emph{alternating bottom-up automaton for unordered unranked
trees with horizontal sub-automata} is defined as 
the union of all classes \(\AUTFun{\H_H}\) such that \(H\) is a horizontal automaton
with alphabet \(\Alphabet\).
\end{defn}


For vertically deterministic automata, filters 
can be applied only to pairs \((d,Q)\) such that \(\card Q\le 1\). Therefore
we will be interested in restricted problems for filters, in which
the state set \(Q\) of all models are either empty or singletons. We
will call the restricted problems of filter \emph{singleton-membership},
\emph{singleton-satisfiability}, \emph{singleton-validity}, etc. It
should be noticed that the singleton-restricted problems are
usually much easier than the general case. For instance,
if \(\Pat=\emptyset\) then singleton-satisfiablity and
singleton validity of filters is in \PTIME. This also
remains true, if only suffixes can be tested by
patterns, i.e., if \(\Pat=\{*\Text{d}\mid d\in \Alphabet^*\}\).

\medskip
\begin{prp}[\AUTA\ Complexities]\label{autacompl}
Given two vertically deterministic \AUTA\ \(A,B\), 
a tree \(t\in \Trees\),
then 
if singleton-satisfiability of \f\ is decidable in time \(O(f(\size \f))\),
whether \(\L(A) = \void\) can be tested in time
\(O(\size A^2 \cdot f(\size A))\),
whether \(t\in \L(A)\) is \NP-complete,
whether \(\L(A)=\Trees\) is \PSPACE-hard,
and
provided that singleton-satisfiability
of a filter \f\
is testable in
polynomial time,
deciding whether \(\L(A)\inter \L(B) = \void\) is \coNP-complete.
\end{prp}

\ifLongPaper

\begin{proof}
\textbf{Emptiness.}
We build the set \(S\incl \Q\) of reachable states.
Initially, we let
\[
S \affectEQ \setst{q}{ \pat\to q,\ \pat \x{ is satisfiable}}
\mstop
\]
Then we iterate, testing for each \(pp'\to q\) whether the rewriting from \(p\)
to \(p'\) can be done using only the states in \(S\); that is,
we determine whether \(p'\) is reachable from \(p\), following the relation
\[
(p,M + \b{d:Q}) \rewh_S (p', M)
\xxx{ if }  
\exists \f: 
(p,\f,p') \in \hRules \ttand (d,Q\inter S) \models \f
\mstop
\]
The iteration step, performed until a fixed point is reached, is:
\[
S \affectEQ S\union \setst{q}{ pp'\to q,\ p' \x{ is reachable from } p \x{ using } S}
\mstop
\]
There remains to see how one actually tests whether \(p'\) is reachable from \(p\) using \(S\).
We build the set \(\bar P\incl P\) of states reachable from \(p\):
initially, \(\bar P = \sset p\). At each step, we execute
\[
\bar P \affectEQ \bar P \union\setst
{p'}
{p \f \to p', \ p\in \bar P, \ \px[\Big]{\f \land \lOr_{q\in S} q} \x{ is satisfiable}}
\]
until a fixed point is reached (failure) or we reach \(p'\) (success).
Thus, the horizontal reachability algorithm 
comprises at most \(\size P\) iteration steps, each performing
\(\size \hRules\) satisfiability tests each doable in \(O(f(\size\hRules + \size \Q))\).
Going back now to the overarching vertical reachability,
its iteration step is performed at most \size \Q\ times,
and executes at most \size \Rules\ horizontal reachability algorithms each time.
Overall, this gives a complexity in 
\(O(\size \Q\cdot\size \Rules \cdot \size P \cdot \size\hRules \cdot f(\size\hRules + \size \Q))\).

This can be simplified considerably, however,
by considering that the horizontal reachability tests are always done on the same
automaton, and that if \(p'\) is reachable from \(p\) using \(S\), then it will remain reachable
using any superset of \(S\). Thus, if one keeps a permanent memory of previous reachability results
for the horizontal automaton, \ie marking states as ``reachable from \(p\)'' for each \(p\),
then over the course of the entire algorithm, each rule in \hRules\ needs to be tested
only once for each state, at most.
This yields a complexity of 
\(O(\size \Q\cdot\size \Rules + \size P \cdot \size\hRules \cdot f(\size\hRules + \size \Q))\).

\textbf{Membership.}
\emph{Lower bound.} Decidability of membership to Parikh images of
the language accepted by 
finite-state word automata over a finite alphabet
is NP-complete \cite{DBLP:conf/lics/KopczynskiT10},
and is easily reduced to \AUTA\ membership of a flat tree,
where the horizontal automaton simulates the original word automaton.
\emph{Upper Bound.} Given a run of \(A\), \ie a tree annotated
by the sets of vertical states and the horizontal states,
it can be verified in polynomial time that the run is correct.

\textbf{Universality.}
Decidability of universality of Parikh images of
the language accepted by 
finite-state word automata over a finite alphabet
is known to be \PSPACE-complete \cite{KopDraft}.

\textbf{Disjointness.}
Decidability of disjointness of Parikh images of
the languages accepted by 
finite-state word automata over a finite alphabet
is known to be coNP-complete \cite{DBLP:conf/lics/KopczynskiT10,KopDraft}.
\end{proof}
\else
\begin{proof}[Proof sketch]
The hardness results follow from known lower bounds, see for instance \cite{DBLP:conf/lics/KopczynskiT10,KopDraft}.
Emptiness follows from a vertical accessibility algorithm where each phase performs
a horizontal accessibility algorithm.
A proof for membership is just a vertical run, with assorted horizontal runs,
checkable in \PTIME, hence the upper bound.
For \(\L(A)\inter \L(B) = \void\),
the polynomial check of \cite{DBLP:conf/lics/KopczynskiT10,KopDraft} can be used in our case by replacing the infinite alphabet \(\Alphabet\) by the pair of rules a labeled datavalue would use in the horizontal automata of \(A\) and \(B\).
To make sure we do not combine two mutually disjunctive rules, we need to make sure in polynomial time that their conjunction is singleton-satisfiable.
\end{proof}
\fi

\section{AUTs with Confluent Horizontal Rewriting}
\label{sec:confluence}

In this section we move towards more tractable classes: 
we define a subclass of vertically deterministic
\AUTAs for which the horizontal automata must be confluent.
Intuitively, that means that, during the horizontal evaluation,
one can choose any available transition in a ``don't care'' manner,
since all possible choices will yield the same result at the end.

The resulting expressive power lies strictly between \CMSO
and \PMSO. For instance, one can test \(\# q = \# q'\), which is 
not in \CMSO, but cannot test \(\# q \< \# q'\), even though this
can be tested in \PMSO. Despite its high expressive power, this model 
has some good static analysis properties.


A horizontal automaton \(H=\tuple{\Alphabet,\hStates,\hRules}\) is called \emph{confluent}
if the \emph{failure-extended horizontal rewriting relation} \(\rewhx\)
is confluent, where \(\rewhx\) is defined as the smallest relation such that
\[
\rewh\ \incl\ \rewhx \ttand
(p,M)\rewhx \bot \xx{if} M\ne \b{} \ttand \nexists p',M' : (p,M)\rewh (p',M')
\mstop
\]
Its \emph{\(p_0\)-confluent descriptor class} \(\Hconf_{H,p_0}\),
for \(p_0\in \hStates\), is the subclass of \(\H_H\)
where the descriptors are limited to \(\sset {p_0}\times\hStates\).
Indeed, having several initial states would be ``cheating'' the confluence.\mk

\begin{defn}[\AUTC]
An \AUTC is a vertically deterministic member of any class \(\AUTFun{\Hconf_{H,p_0}}\),
where \(H=\tuple{\Alphabet,\hStates,\hRules}\) is a confluent horizontal automaton, and \(p_0\in\hStates\).
\end{defn}\mk

\begin{prp}[\AUTC\ Closure Properties]\label{autc-closure}
\AUTCs are neither closed under union nor complement.
\end{prp}
\ifLongPaper\begin{proof}
This property comes from the fact that horizontal confluent automata on a finite alphabet are not stable under union or negation.
The language \(\#a=\#b\) is recognized by a confluent automaton.

\textbf{Union:} \(\#a=\#b\lor\#a=\#c\) cannot be accepted.
To prove it, consider the multisets \(M_n\) with \(n\) letters \(a\), \(n\) letters \(b\), no \(c\).
They should be accepted, but all in different states than any other \(M_{n'}\), as adding one letter \(b\) and \(n\) letters \(c\) would still work for \(M_n\), but not for \(M_{n'}\).

\textbf{Negation:} \(\#a\neq\#b\) cannot be accepted.
To prove it, consider this time the multisets \(M_n\) with \(n\) letters \(a\), no \(b\).
They should be accepted, but all in different states than any other 
\(M_{n'}\), as adding \(n\) letters \(b\) would not work for \(M_n\), but should for \(M_{n'}\).
\end{proof}
\else
\begin{proof}
\(\#\Text{a}=\#\Text{b}\) and \(\#\Text{a}=\#\Text{c}\) are recognizable by a confluent automaton, \(\#\Text{a}=\#\Text{b}\lor\#\Text{a}=\#\Text{c}\) is not: the class is not closed under union.
\(\#\Text{a}=\#\Text{b}\) is recognizable by a confluent automaton, \(\#\Text{a}\neq\#\Text{b}\) is not: the class is not closed under complement.
\end{proof}
\fi

\medskip
\begin{prp}[\AUTC\ Membership]\label{autc-membership}
If singleton-membership of filters is in \PTime, then one can
decide for any  \AUTC\ \(A\) and tree \(t\) whether \(t\in\L(A)\) in polynomial time.
\end{prp}
\ifLongPaper
\begin{proof}
Since our horizontal automaton is confluent, the greedy strategy works to read arities:
We start reading \(M\) in state \(p\).
For every data value in the arity, we try every transition (time \(|M|\times |A|\times \) membership on the filter).
If we cannot read a data value, the run fails. While we can, we pick any data value we can read.
Since the automaton is confluent, this strategy will end in the same state - or fail - no matter which choice is made at each step.
\end{proof}
\else
\begin{proof}
Since the horizontal automaton is confluent, the greedy strategy of reading a data value 
whenever we can always gives the proper result.
We make such a test for each node of the input tree in a bottom-up manner.
\end{proof}
\fi

\begin{prp}[\AUTC\ Emptiness]\label{autc-emptiness}
If singleton-satisfiability of filters is in \PTime, then
it is decidable in polynomial time for an \AUTC\ \(A\), whether 
\(\L(A)=\void\).
\end{prp}
\begin{proof}
This is a particular case of Prop \ref{autacompl}.
\end{proof}

\begin{prp}[\AUTC\ Universality]\label{autc-universality}
If the singleton-validity of filters is in \PTime,
then it is decidable in polynomial time for any 
\AUTC\ \(A\) whether \(\L(A)=\Trees\).
\end{prp}
\ifLongPaper
\begin{proof}
First, we map all accessible vertical states, and all accessible horizontal states.
For the automaton to be complete,
all trees must be accepted in a final state, which means all vertical accessible states must be final,
all trees must be accepted in some state, which means all horizontal accessible states must be final,
but also that all horizontal accessible states must be able to read any pair \((d,\{q\})\), where \(q\) accessible
(this means that the union of the filters for all rules exiting a state is universal).
All these test can be made in polynomial time.
It is easy to prove that these condition are not only necessary, but also sufficient for the automaton to be complete.
\end{proof}
\else
\begin{proof}
We check that all vertical (resp. horizontal) accessible states are accepting and can read any possible arity (resp. labeled data value).
\end{proof}
\fi
\begin{prp}[\AUTC\ Disjointness]\label{autc_disjoint}
If the singleton-satisfiability of the conjunction of two filters is in \PTime, 
then deciding for two \AUTC\ \(A_1\), \(A_2\),  whether \(L(A_1)\cap L(A_2)=\emptyset\) is \coNP-complete.
\end{prp}
\ifLongPaper

\begin{proof}
\textbf{coNP-Hard:} As usual, coNP-hardness is directly inherited from the same problem on horizontal automata:
given two confluent horizontal automata \(H_1\) and \(H_2\) on the same set of states \(\Q\),
and two descriptors \((p_1,p'_1)\) of \(H_1\) and \((p_2,p'_2)\) of \(H_2\),
it is NP-hard to decide whether there exist a multiset \(M\) such that \(M\models (p_1,p'_1)\) and \(M\models (p_2,p'_2)\),
even if the satisfiability of the conjunction of two filters is decidable in polynomial time.

We reduce the problem of 3-coloring of an undirected graph \(G=\left\{V,E\right\}\) with colors \(C=\left\{x, y, z\right\}\).
Our alphabet is the set of 
\pointl1
pairs \((v,c)\in V\times C\), that represents that the node \(v\) has color \(c\),
and 
\pointl2 the tuples \((v,c,v',c')\), if \(c\neq c'\) and \(v,v'\) are neighbors.
In particular, there are no tuples of the form \((v,c,v',c)\).
We shall consider arities that represent a coloring:
for each \(v\in V\), exactly one letter \((v,c)\) to color the node \(v\) with color \(c\),
then for every \(v'\) neighbor of \(v\) , and \(v'\) of color \(c'\) the edges \((v,c,v',c')\) and \((v',c',v,c)\) appear once.
It is easy to see that such an arity exists if and only if there is a 3-coloring of \(G\)
We now want to create \(H_1\) and \(H_2\) two confluent automata whose intersection will be exactly these arities:

We number the states of \(V\) as \(\enum v1n\).
The automaton \(H_1\) will check that each node \(v\) has exactly one letter \((v,c)\) to color the node \(v\) with color \(c\),
and for every neighbor \(v'\) an edge \((v,c,v',c')\).
\begin{itemize}
\item For each node \(v\in V\), there is a state \(p_{v}\),and a state \(p_{v,c}\) for each \(c\in C\).
We have the transitions \(p_v,(v,c)\rightarrow p_{v,c}\).
This checks \(v\) is colored.
\item We note \(N_v=\left\{n_1\ldots n_k\right\}\) the set of neighbors of \(v\).
We have \(k+1\) states \(p_{v,c,n_j}\), where \(p_{v,c,n_1}\) is \(p_{v,c}\) and \(p_{v,c,k+1}\) is \(p_{v'}\), the starting state for the next node in \(v\).
These states check the edge from \(v\) to \(n_j\).
We have the rules \(p_{v,c,n_j},(v,c,n_j,c')\rightarrow p_{v,c,n_{j+1}}\).
\item For the first node \(v_0\), \(p_{v_0}\) is initial.
After the last neighbor of the last node of \(V\), we have a final state \(p_f\).
\end{itemize}
To prove the confluence, we note that any letter can only be read in one state and that the automaton contains no loop,
which means if at one point we have a choice to read two different letters, then no matter the choice, the run will fail.

Automaton \(H_2\) will check that each node \(v\) has exactly one letter \((v,c)\) to color the node \(v\) with color \(c\),
and for every neighbor \(v'\) an edge \((v',c',v,c)\).
It is the same as \(H_2\) except the rules \(p_{v,c,n_j},(v,c,n_j,c')\rightarrow p_{v,c,n_{j+1}}\) becomes \(p_{v,c,n_j},(n_j,c',v,c)\rightarrow p_{v,c,n_{j+1}}\)
We can see that that \(\L(A_1)\inter \L(A_2)\) is exactly the arities that represents colorations of \(G\),
hence \(L(A_1)\cap L(A_2)=\emptyset\) if and only if there is no 3-coloring of \(G\).

\textbf{coNP:} This problem is already in \coNP for the non-confluent case, which makes it \coNP-complete in this particular case.
\end{proof}
\else
\begin{proof}
The \coNP-hardness is inherited from the same problem on horizontal automata:
given two confluent horizontal automata \(H_1\) and \(H_2\) on the same set of states \(\Q\),
and two descriptors \((p_1,p'_1)\) of \(H_1\) and \((p_2,p'_2)\) of \(H_2\),
decide whether there exist a multiset \(M\) on singletons such that \(M\models (p_1,p'_1)\) and \(M\models (p_2,p'_2)\).
This result is a reduction of 3-coloring a graph:
we encode successful colorings
as a Parikh language in the intersection of \(L(p_1,p'_1)\cap L(p_2,p'_2)\).
The problem is already in \coNP for \AUTA by Proposition \ref{autacompl}, so it is also in \coNP 
for the more restricted class \AUTC.
\end{proof}
\fi

\begin{prp}[\AUTC\ Inclusion]\label{autc_include}
If the singleton-satisfiability of filters is in \NP, then
deciding for \AUTC\ \(A_1\) and \(A_2\) whether \(L(A_1)\subseteq L(A_2)\) is \coNP-complete.
\end{prp}

%
%
%
%
\ifLongPaper

\begin{proof}
\textbf{coNP-Hard:} As usual, coNP-hardness is directly inherited from the same problem on horizontal automata:
given two confluent horizontal automata \(H_1\) and \(H_2\) on the same set of states \(\Q\),
and two descriptors \((p_1,p'_1)\) of \(H_1\) and \((p_2,p'_2)\) of \(H_2\),
it is NP-hard to decide whether there exist a multiset \(M\) such that \(M\models (p_1,p'_1)\) but \(M\not\models (p_2,p'_2)\),
even if the satisfiability of the conjunction of two filters is decidable in polynomial time.

This proof is an adaptation of the proof of Prop. \ref{autc_disjoint}:
\(H_1\) remains the same, but \(H_2\) will check something slightly different.
It will check that each node \(v\) has exactly one letter \((v,c)\) to color the node \(v\) with color \(c\),
and for every neighbor \(v'\) an edge \((v',c',v,c)\),
except for at least one node \(v\) colored in \(c\) but with a miscolored edge \((v',c',v,c''\neq c)\).
\begin{itemize}
\item For each node \(v\in V\), there is a state \(p_{v}\),and a state \(p_{v,c}\) for each \(c\in C\).
We have the transitions \(p_v,(v,c)\rightarrow p_{v,c}\).
This checks \(v\) is colored.
\item We note \(N_v=\left\{n_1\ldots n_k\right\}\) the set of neighbors of \(v\).
We have \(k+1\) states \(p_{v,c,n_j}\), where \(p_{v,c,n_1}\) is \(p_{v,c}\) and \(p_{v,c,k+1}\) is \(p_{v'}\), the starting state for the next node in \(v\).
These states check the edge from \(v\) to \(n_j\).
We have the rules \(p_{v,c,n_j},(n_j,c',v,c)\rightarrow p_{v,c,n_{j+1}}\).
\item For the first node \(v_0\), \(p_{v_0}\) is initial.
\end{itemize}
To this automaton, we add a copy where the run goes if it detects the error it was looking for:
\begin{itemize}
\item For each node \(v\in V\), there is a state \(p'_{v}\),and a state \(p'_{v,c}\) for each \(c\in C\).
We have the transitions \(p'_v,(v,c)\rightarrow p'_{v,c}\).
\item We note \(N_v=\left\{n_1\ldots n_k\right\}\) the set of neighbors of \(v\).
We have \(k+1\) states \(p'_{v,c,n_j}\), where \(p'_{v,c,n_1}\) is \(p'_{v,c}\) and \(p'_{v,c,k+1}\) is \(p'_{v'}\), the starting state for the next node in \(v\).
We have the rules \(p'_{v,c,n_j},(n_j,c',v,c'')\rightarrow p'_{v,c,n_{j+1}}\)
(this allows for correct edges or errors after the first one to not fail the run).
\item We travel from the part "without error" to the part "with error" with rules \(p_{v,c,n_j},(n_j,c'\neq c,v,c''\neq c)\rightarrow p'_{v,c,n_{j+1}}\).
\item After the last neighbor of the last node of \(V\), we have a final state \(p'_f\).
\end{itemize}
To prove the confluence, we note that any letter can only be read in one state and its copy in the error automaton,
and that a run cannot visit both,
which means if at one point we have a choice to read two different letters, then no matter the choice, the run will fail.

The \coNP check on horizontal automata is proper to the confluent restriction, as the problem is \PSPACE-hard in the general case. Since membership is polynomial we just have to ensure that there exists counter-example of polynomial size in case inclusion does not hold.
Let \(H_1\) be a confluent automaton, \(p_0\) an "initial" state, \(\enum p1n\) "final" states.
A \emph{minimal acceptor} \(M\) of \(p_i\) is a multiset such that there is a state \(p_j\) (\(j\neq 0\)) such that \(M \models (p_i,p_j)\),
but for every \(M'\subseteq M\) there is no final state \(p\) such that \(M' \models (p_i,p)\).
Since we consider confluent automata, these minimal acceptors describe a greedy strategy for accepting a multiset:
we can partition \(M\) into minimal acceptors \(M_0\) from \(p_0\) to a \(p_{i_1}\), then \(M_1\) from \(p_{i_1}\) to a \(p_{i_2}\) \(\ldots\) until \(M\) is entirely read.
Since \(H_1\) is confluent, this method works if and only if \(M\) goes from \(p_0\) to a final state \(p\).
From this, we now consider a second confluent automaton \(H_2\), with its initial state \(p'_0\) and its "final" states \(\enum {p'}1m\).
We try to read \(M\) in \(H_2\) the same way: \(M_0\) first, then \(M_1\)\ldots\ 
If at any step \(M_k\) we do not end up in a final state in \(H_2\), then \(M_0+\dots +M_k\) is a counter-example for the inclusion.
If \(M_0 + \dots + M_k\) ends up in a final state of \(H_2\) then we get a pair of final states \((p_{i_k},p'_{j_k})\).
By a pumping argument, we can get rid of loops and need a less-than-quadratic number of minimal acceptors to reach a counter-example.
Each minimal acceptor cannot be bigger than the number of states in \(H_1\).
Since there is a counter-example of polynomial size, and the membership problem is polynomial, we have a polynomial check.

We use this in the vertical automaton:
We consider two \AUTC\ \(A_1\), \(A_2\).
To put their labeling on the same automaton, we make sure that  \(H_1\) and \(H_2\) now test arities labeled on pairs of \(\Q_1\times\Q_2\).
We can nondeterministically guess a set of accessible pairs by using Prop \ref{autc_disjoint}.
From there, we nondeterministically guess
a counter-example of arity labeled on these accessible pairs that leads to a final state in \(A_1\) but not in \(A_2\).
\end{proof}
\else
\begin{proof}
Again, \coNP-hardness is inherited from the analogous problem on horizontal automata:
given two \AUTCs \(H_1\) and \(H_2\) on the same set of states \(\Q\),
and two descriptors \((p_1,p'_1)\) of \(H_1\) and \((p_2,p'_2)\) of \(H_2\),
decide whether there exist a multiset \(M\) on singletons such that \(M\models (p_1,p'_1)\) but \(M\not\models (p_2,p'_2)\).
This result
is a reduction of 3-coloring a graph: we encode successful colorings
as a Parikh language recognized by \(L(p_1,p'_1)\backslash L(p_2,p'_2)\), with \(H_1\) and \(H_2\) confluent horizontal automata.

The \coNP check on horizontal automata is proper to the confluent restriction, as the problem is \PSPACE-hard in the general case.
Since membership is polynomial we just have to ensure that there exists a counter-example of polynomial size in case inclusion does not hold.
Let \(H_1\) be a confluent automaton, \(p_0\) an "initial" state, \(\enum p1n\) "final" states.
A \emph{minimal acceptor} \(M\) of \(p_i\) is a multiset such that there is a state \(p_j\) (\(j\neq 0\)) such that \(M \models (p_i,p_j)\),
but for every \(M'\subseteq M\) there is no final state \(p\) such that \(M' \models (p_i,p)\).
Since we consider confluent automata, these minimal acceptors describe a greedy strategy for accepting a multiset:
we can partition \(M\) into minimal acceptors \(M_0\) from \(p_0\) to a \(p_{i_1}\), then \(M_1\) from \(p_{i_1}\) to a \(p_{i_2}\) \(\ldots\) until \(M\) is entirely read.
Since \(H_1\) is confluent, this method works if and only if \(M\) goes from \(p_0\) to a final state \(p\).
From this, we now consider a second confluent automaton \(H_2\), with its initial state \(p'_0\) and its "final" states \(\enum {p'}1m\).
We try to read \(M\) in \(H_2\) the same way: \(M_0\) first, then \(M_1\)\ldots\ 
If at any step \(M_k\) we do not end up in a final state in \(H_2\), then \(M_0+\dots +M_k\) is a counter-example for the inclusion.
If \(M_0 + \dots + M_k\) ends up in a final state of \(H_2\) then we get a pair of final states \((p_{i_k},p'_{j_k})\).
By a pumping argument, we can get rid of loops and need a less-than-quadratic number of minimal acceptors to reach a counter-example.
Each minimal acceptor cannot be bigger than the number of states in \(H_1\).
Since there is a counter-example of polynomial size, and the membership problem is polynomial, we have a polynomial check.

We use this in the vertical automaton:
We consider two \AUTC\ \(A_1\), \(A_2\).
To put their labeling on the same automaton, we make sure that  \(H_1\) and \(H_2\) now test arities labeled on pairs of \(\Q_1\times\Q_2\).
We can nondeterministically guess a set of accessible pairs by using Prop \ref{autc_disjoint}.
From there, we nondeterministically guess
a counter-example of arity labeled on these accessible pairs that leads to a final state in \(A_1\) but not in \(A_2\).
\end{proof}
\fi

\section{AUTs with Ordered Horizontal Rewriting}
\label{sec:fixed order}

We now introduce a subclass whose expressive power id even more restricted than that of \AUTC.
In this class, the filters are required to be disjoint
-- or made to be so beforehand at quadratic cost -- and
are linearly ordered, the order itself being a parameter of the class.
Each rule has its own horizontal automaton, restricted to reading the arity following this order.
Compared to the confluent case, \(\#a = \#b\) for instance is no longer expressible, 
but \(\#a \equiv \#b \mod k\) still is.

The filters in a given horizontal automaton being disjoint, we
let \(\gS = \setenum \f1n\) be the chosen finite alphabet of filters, 
and view an arity
\(\b{d_1:Q_1,\ldots,d_n:Q_m}\) as 
\(\b{\enum \gp1m}\), where \(\gp_i\) is the unique \(\f \in \gS\) such that
\((d_i,Q_i) \models \f\); this is undefined if there is no such  \(\f\).
Thus we see arities as Parikh images of words on the finite alphabet \gS,
and horizontal automata as -- deterministic -- finite state automata on \gS.

\emph{Deterministic finite automata} (DFA) are defined as usual as tuples 
\(\dfa=\tuple{\gS,P,\ini p, \fin P, \gd}\), where \(\gd : P \times \gS \to P\).
We write a transition simply \(p\f\to p'\).
The word language of a DFA \(\dfa\) is written \(\L(\dfa)\), and
its \emph{Parikh language} is the Parikh image of its word language,
\ie \(\P(\dfa)=\setst{\b{\enum\gp1m}}{\enumop\gp1m{} \in \L(\dfa)}\).
Given a linear order \(\prec\) on \gS\ (or on \(\Filters \sups \gS\))
such that \enumop\f1n\prec, the \(\prec\)-ordered language of 
\(\dfa\) is \(\L_\prec(\dfa) = \L(\dfa) \inter \f_1^* \dots \f_n^*\),
and we also let \(\P_\prec(\dfa)\) be the Parikh image of  \(\L_\prec(\dfa)\).

We define the corresponding horizontal descriptor class
\tuple{\DFAa,\models,\size\cdot,0},
with \(\DFAa\) being the set of DFA on a set \(\gS\incl \Filters\) of mutually disjoint filters,
\(\size\dfa\) being the usual
size for DFA, and the following satisfaction relation:
\[
\b{d_1:Q_1,\ldots,d_m:Q_m} \models \dfa
\xx{iff}
\forall i=1..m,\ 
\exists \gp_i \in \gS :
d_i:Q_i \models \gp_i \lland
\b{\enum\gp1m} \in \P_\prec(\dfa)
\mstop
\]

\begin{defn}[\AUTO]
An 
\AUTO is a is vertically deterministic member of \(\AUTFun{\DFAa}\).
\end{defn}\mk

\bk

\begin{prp}[Reordering]\label{auto-reordering}
For any \AUTO\ A, and any total filter order \(\prec'\),
one can construct an \AUTO[\prec'] \(A'\) equivalent to \(A\)
in time 
\(O(2^{{\size A} \cdot \card \gS })\).
\end{prp}
\ifLongPaper
\begin{proof}
\(A'\) has the same structure as \(A\), except that
each rule \(\dfa\to q\) is replaced by a rule \(\dfa'\to q\),
where \(\dfa,\dfa'\) are such that \(\P_\prec(\dfa)= \P_{\prec'}(\dfa')\);
given that, it is immediate that \(A\) and \(A'\) are equivalent.
Thus it is sufficient to show how to build \(\dfa'\), given \(\dfa=\tuple{\gS,P,\ini p, \fin P, \gd}\), such that
\(\L_\prec(\dfa)= \L_{\prec'}(\dfa')\) -- which implies the equality of Parikh
images. 
Let \(\gS=\setenum a1n\) be the finite alphabet of both \dfa\ and \(\dfa'\),
such that \(\enumop a1n\prec\) and 
\(a_{\gs(1)} \prec' \dots \prec' a_{\gs(n)}\).
We partition \dfa\ into \(n\) components \enum C1n, 
which are ``DFA'' with potentially several initial states,
each corresponding to
the treatment of a letter: 
\[
C_i = \tuple{\gS,P,I_i, F_i, \gd \inter P\times \sset{a_i}\times P}
\mcom
\]
where \(I_1 = \sset{\ini p}\), \(I_i = \setst{p}{\exists p': p' a_{i-1} \to p}\), for \(i\>2\), and
and \(F_n = \fin P\), \(F_i = \setst{p}{\exists p': p a_{i+1} \to p'}\), for \(i<n\).
With this, we have a compartmentalised vision of the test effected by \dfa\
on a word \(w \in a_1^* \dots a_n^*\):
we begin in the initial state of \(C_1\), and test a property
on the \(a_1^*\) prefix of \(w\); depending on the outcome,
we end up into different ``final'' states of \(F_1\),
which are also ``initial'' states of \(I_2\), and so on
until we reach a final state of \(F_n\), which concludes the test.

In this vision, each final state of \(F_i\) encodes a specific property
on the \(a_i^*\) part of \(w\) 
and makes accessible certain other properties
for the next component -- and the ones after that by transitivity.
We assimilate the final states and their properties,
assuming wlog. that no state of \(F_i\) is reachable by two distinct
states of \(I_i\).
Thus the tests realised can be represented by an edge-labelled tree,
where, coming from an edge labelled by \(p\in F_{i-1}\), 
the edges of the new node are decorated by those \(F_i\)
which are reachable from \(p\) -- which also belongs to \(I_i\).
The new nodes thus created are of course populated recursively;
initially all states of \(F_1\) are considered reachable.

Each path on that tree -- call it \(T\) -- describes a conjunction of
properties \enum\gps1n, where \(\gps_i\) pertains to
the \(a_i^*\) factors of \(w\), the language of \dfa\ being
the disjunction of those conjunctions of properties.
Note that this formulation is not dependent upon the
order in which the \(a_i^*\) factors of \(w\) are processed.
To build \(\dfa'\), we need only ensure that the same construction,
with \(\prec'\) this time, yields another tree with the same paths.
Taking a state-centric vision, \(T\) also provides us with a blueprint
for rebuilding \dfa\ from the \(C_i\).
Instead, we shall first rearrange the tree's levels according to the new order
\(a_{\gs(1)} \prec' \dots \prec' a_{\gs(n)}\), preserving the paths,
and build \(\dfa'\) from that new tree \(T'\).

Although it is not always needed, it is convenient to make it so
all \(C_i\) are true DFA, with a single initial state, 
by computing the product of all the variants of \(C_i\) with a single
initial state \(p\in I_i\).
The new final states may now contain several states of \(F_i\),
and thus encode conjunctions of properties; \(T'\) is updated
to reflect that, by duplicating every \(\gp: U\), where \gp\ is a property 
and \(U\) the subtree beneath it, into \(\set{\gp,\dots} : U\) for all subsets of
properties containing \gp\ satisfied simultaneously in the new states.
Overall, this is done at up to exponential cost, because of the products.
(Note that the products are not always needed later, so there is room here for optimisation).

All that is left to do is to follow the blueprint provided by \(T'\)
to combine our new \(C_i\) into \(\dfa'\).
We begin by \(C_{\gs(1)}\), setting its initial state as the initial state of \(\dfa'\),
then each of the states of \(F_{\gs(1)}\) also becomes the initial state of
-- an new instance of --
\(C_{\gs(2)}\) and so on until the end, thus forming a tree of automata of the same shape as \(T'\).
There remains to mark the final states of \(\dfa'\), which are,
for each instance of \(C_{\gs(n)}\), the final states which are also edge labels of the corresponding
node of \(T'\).

This construction brings another exponential, because in the worst case there is a copy of a \(C_i\)
for each node of \(T'\).
\end{proof}
\else\mk
In practice, a direct algorithm considering the ``decision trees''
underlying the horizontal automata, testing each \f\ separately in the
order $\prec$, and reordering the decisions according to $\prec'$,
generally avoids explosive size increase.  Its asymptotic bound is currently worse,
though: \(O(2^{2{\size A}\cdot\card \gS })\).\mk
\fi

\begin{prp}[\CMSO-Equivalence]
For any total order \(\prec\), \AUTO\ has exactly the same expressive 
power as \CMSO.
\end{prp}
\begin{proof}[Proof sketch]
It is obvious that \(\DFAa\) can encode counting constraints,
as they can encode 
\(\#a \< k\) and
\(\#a = k \mod  n\),
and are closed under Boolean operations.
Conversely, \(\DFAa\) can be seen as a succession of components
dealing with \(\f_i^*\)-factors
-- as for reordering, --
each of which can be put into Chrobak normal form \cite{gawrychowski2011chrobak},
and can hence be expressed as a disjunction of modulos.
\end{proof}

\begin{prp}[\AUTO\ is Easy]
Given an order \(\prec\) on filters,
the membership, emptiness, universality, disjointness, equivalence, and inclusion
decision problems for vertically deterministic 
\AUTO\ are all polynomial, provided that the corresponding singleton problems for filters are.
\end{prp}
\begin{proof}
This follows from the same results for DFA.
\end{proof}

\begin{table}
\centering
\begin{tabular}{r>{\ \ }llllll} 
& {\AUTP}& {\AUTA}& {\AUTC}& {\AUTO}
\\ \midrule Characterisation:
& \PMSO          & \PMSO          &  \CMSO \(<\cdot<\) \PMSO            & \CMSO
\\ \(t\in \L(A)\) ? 
& in \PTIME        & \NP-complete   & in \PTIME        & in  \PTIME
\\ \(\L(A) = \void\) ? 
& \PSPACE-hard     & in \PTIME        & in \PTIME        & in \PTIME
\\ \(\L(A)\inter \L(B) = \void\) ? 
& \PSPACE-hard     & \coNP-complete & \coNP-complete     & in \PTIME
\\ 
\(\L(A)=\Trees\) ? 
& \PSPACE-hard  & \PSPACE-hard  &  in \PTIME        & in \PTIME
\\
\(\L(A) = \L(B)\) ? 
& \PSPACE-hard  & \PSPACE-hard  &  in \coNP            & in \PTIME
\\ \(\L(A) \incl \L(B)\) ? 
& \PSPACE-hard  & \PSPACE-hard  &  \coNP-complete    & in \PTIME
\\ \bottomrule
\end{tabular}
\caption{Overview of the complexity results for vertically
  deterministic \AUTs with various assumptions on patterns or filters.}
\label{table}
\end{table}

\section{Conclusion and Future Work}

We have introduced a very general setting for bottom-up automata
on unranked unordered trees on infinite alphabets, which captures
the usual notions of alternation 
and 
determinism with respect to the \emph{vertical} -- bottom-up -- structure of automata,
and is parametrized by the modality of \emph{horizontal} evaluation.
We have shown that this model, with Presbuger formul\ae{} or Parikh-like automata,
captures \PMSO, with complexity trade-offs between membership and emptiness.
Searching for classes suitable both for querying and static analysis,
we then examined two notions of \emph{horizontal} determinism:
confluence and fixed-orderedness, the latter yielding the same expressive power as \CMSO,
and the former a strict intermediate between \CMSO\ and \PMSO.
Our complexity results are summarized in Table \pref{table}.

To extend this work, we intend to explore more powerful variants where filters support data joins,
and to generalize the approach to tree transducers, with applications
to static verification of scripts, 
some subclasses 
of which 
can be seen as transducers on filesystem trees.

\bibliographystyle{eptcs}
\bibliography{bib}

\end{document}